\documentclass[9pt]{article}  \usepackage{times}
\usepackage{graphicx, textgreek}

\usepackage{color}

\usepackage[round,numbers,sort&compress]{natbib} 

\usepackage{amsmath}
\usepackage{nccmath}   
\usepackage{amssymb}
\usepackage{stackrel}
\usepackage{chemarrow}
\usepackage{graphicx}
\usepackage{textgreek}

\usepackage{booktabs}
\usepackage{dcolumn}
\usepackage{rotating}
\usepackage{multirow}

\begin{document}

	\twocolumn[{\LARGE \textbf{The effect of stretching on nerve excitability\\*[0.2cm]}}
	{\large Thomas Heimburg$^\ast$\\*[0.1cm]
		{\small Niels Bohr Institute, University of Copenhagen, Blegdamsvej 17, 2100 Copenhagen \O, Denmark}\\*[-0.1cm]
		
		{\normalsize \textbf{ABSTRACT}\hspace{0.5cm} Nerves are frequently stretched during movement. We investigate here the effect of stretch on nerve excitability within the framework of the soliton theory. This thermodynamic theory for nerve pulse propagation relies on the presence of a melting transition in the nerve membrane. In this transition, the area of the nerve membrane and the nerve thickness change. It depends on thermodynamic variables including temperature, the chemical potentials of anesthetics and on hydrostatic pressure. A further variable relevant for movement science is the the stretching of nerves, i.e., a tension in the nerve caused by muscle contraction, the bending of joints and the pulling on extremities. We show here that the soliton theory predicts a decrease in nerve excitability upon stretching. This becomes evident in a reduction of the amplitude of compound action potentials and in the suppression of reflexes. We compare these predictions with medical findings.			\\*[0.3cm] }}
	\noindent\footnotesize{\textbf{Keywords:} stretching, reflexes, solitons\\*[0.1cm]}
	\noindent\footnotesize {$^{\ast}$corresponding author, theimbu@nbi.ku.dk. }\\
	\vspace{0.3cm}
	]

	\normalsize

\section{Introduction}
\label{introduction}

Nerves increase their thickness \cite{Iwasa1980a, Iwasa1980b, Tasaki1989, Akkin2007, Kim2007, GonzalezPerez2016} and reduce their length \cite{Wilke1912a, Wilke1912b, Bryant1955, Tasaki1980b, Tasaki1982a, Nakaye1985, Tasaki1989} during the action potential. Taking the apparent mechanical features of the action potential as an experimental fact, it is to be expected that the application of mechanical forces on nerves will influence the propagation of the nerve pulse. In this article we will focus on nerve stretching. Stretching of nerves can occur during bending of joints or by forceful elongation of extremities and their muscles. There exist in fact numerous indications that upon stretching nerves become less excitable and that the amplitude of compound action potentials is reduced \cite{Wall1992, Kwan2009, Guissard2001, Guissard2006, Budini2016}.

In this article we wish to discuss this phenomenon in the context of the soliton theory for nerves \cite{Heimburg2005c}, which focuses on the mechanical and thermodynamic aspects of nerve pulse conduction. The soliton theory considers the nerve membrane as a thermodynamic ensemble. Its physical state depends on intensive variables such as pressure, temperature and membrane tension. In particular, the nerve membrane displays a melting transition between an ordered and a disordered state of the membrane lipids, in which all of the extensive variables change. Such transitions have been found in numerous biological membranes (including neural membranes) some 10--15 degrees below physiological temperature \cite{Heimburg2007a, Muzic2019, Faerber2022}, see Fig. \ref{figure1}. In the melting transition, the order of the membrane lipids changes from an ordered solid (or gel) low-enthalpy state to a disordered liquid (or fluid) high enthalpy state. The exact value of the melting temperature depends on the intensive variables pressure, membrane tension, voltage, pH, the concentration of anesthetics and membrane proteins. Importantly, in the transition the magnitude of the susceptibilities is generally higher. For instance, both heat capacity, volume and area compressibility reach maxima \cite{Heimburg1998}.

Due to the nonlinear changes in the lateral compressibility close to the melting transition, the surprising consequence is the possibility of localized density changes (called solitons) in the nerve membrane that share many properties with the nerve pulse \cite{Heimburg2005c, Lautrup2011}. In this theory, the exact melting temperature is important. Therefore, organisms undergo some effort to adapt to changes in environmental conditions in order to keep the distance of physiological temperature and melting temperature constant. Following a decrease in ambient temperature, the membranes of \emph{E. coli} membranes adapt their lipid composition such that the melting temperature of membranes is also lowered \cite{Muzic2019}. Changes in the intensive variables can change the transition temperature. Anesthetics lower the melting temperature and thereby render the nerve membrane less excitable \cite{Heimburg2007c}. In contract, pressure increases melting temperatures and thereby can compensate for the effect of anesthetics, leading to the well-known phenomenon of the pressure-reversal of anesthesia \cite{Johnson1950, Heimburg2007c}. Similar dependencies of melting transitions are known for pH changes and the changes in concentrations of the salts of Na$^+$, K$^+$ \cite{Trauble1976} and even stronger by salts of Li$^+$ and divalent ions such as Ca$^{2+}$ \cite{Binder2002}.

In the present paper we wish to focus on another variable that might be of medical relevance, which is the stretching of nerves. When nerves are stretched, for example when a joint is bent or muscles are contracted, this increases the surface area of the nerves. Tension is generated in the nerve membrane, which has an influence on the melting temperature of membranes. Compressing a membrane (lateral pressure) leads to a higher melting temperature while a stretching of the membrane (tension) leads to lower melting temperatures. It was shown in \cite{Wang2018} that the lowering of the membrane transition temperature caused by anesthetics leads to an increase of the stimulation threshold, in agreement with clinical experiments. Along these lines, we show here that lateral tension also generates an increase of the stimulation threshold for nerve pulse, or a lowering of the action potential amplitude at constant stimulus.

\section{Theory}
\label{theory}


\subsection{Nerve stretching}
\label{nervestretching}

Let us consider a membrane cylinder of radius $r_0$ and length $x_0$. It's volume is
\begin{equation}
V_0=x_0\cdot \pi r_0^2
\label{eq:stretch_01}
\end{equation}
and its membrane area is
\begin{equation}
A_0=x_0\cdot 2\pi r_0 =\sqrt{4\pi x_0 V_0} \;.
\label{eq:stretch_02}
\end{equation}

Let us now stretch the nerve to a length $x=x_0+\Delta x$. We assume that the aqueous medium inside the cylinder is incompressible and that therefore the nerve volume $V=V_0$ is constant. This implies that the membrane area increases upon increasing the length $x$. The area is given by
\begin{equation}
A=A_0\cdot\sqrt{\frac{x}{x_0}}\approx A_0 \left(1+\frac{1}{2}\frac{\Delta x}{x_0}\right)
\label{eq:stretch_03}
\end{equation}
for small $\Delta x/x_0$. For $x>x_0$, one finds a tension $\Pi$ in the membrane. The elastic energy is given by
\begin{equation}
\Delta E=\frac{1}{2 }K_T^A A_0\left(\frac{\Delta A}{A_0}\right)^2 \;,
\label{eq:stretch_04}
\end{equation}
where $K_T^A$ is the lateral compression modulus. The lateral pressure in the membrane is
\begin{equation}
\Pi =-\frac{\partial \Delta E}{\partial A} = -K_T^A \frac{\Delta A}{A_0}=-\frac{1}{2} K_T^A \frac{\Delta x}{x_0}=-\frac{1}{2 \kappa_T^A}\frac{\Delta x}{x_0} \;,
\label{eq:stretch_05}
\end{equation}
where $\kappa_T^A=1/K_T^A$ is the lateral compressibility. Biological membranes exist in the fluid membrane phase. For dipalmitoyl phosphatidylcholine (DPPC) membranes at 50$^\circ$C, $\kappa_T^A=6.9$ m\slash N \cite{Heimburg1998}. Since DPPC is an abundant lipid in biological membranes, we will assume in the following that the lateral compressibility of fluid biological membranes at physiological temperatures is similar. Thus
\begin{equation}
\Pi =-\frac{1}{2 \cdot 6.9 \frac{\mbox{m}}{\mbox{N}}}\frac{\Delta x}{x_0} \;.
\label{eq:stretch_06}
\end{equation}
In the following, we use the term tension for negative lateral pressures. As an example, if $\Delta x/x_0=0.2$, one finds a lateral tension of $-\Pi=0.0145$ N\slash m. For a neuron with a radius of $r_0=10$ \textmu m, this corresponds to a force of -910 nN.


\subsection{Changes in melting temperature induced by nerve stretching}
\label{changesinmeltingtemperatureinducedbynervestretching}

\begin{figure}[htbp]
\centering
\includegraphics[width=8cm]{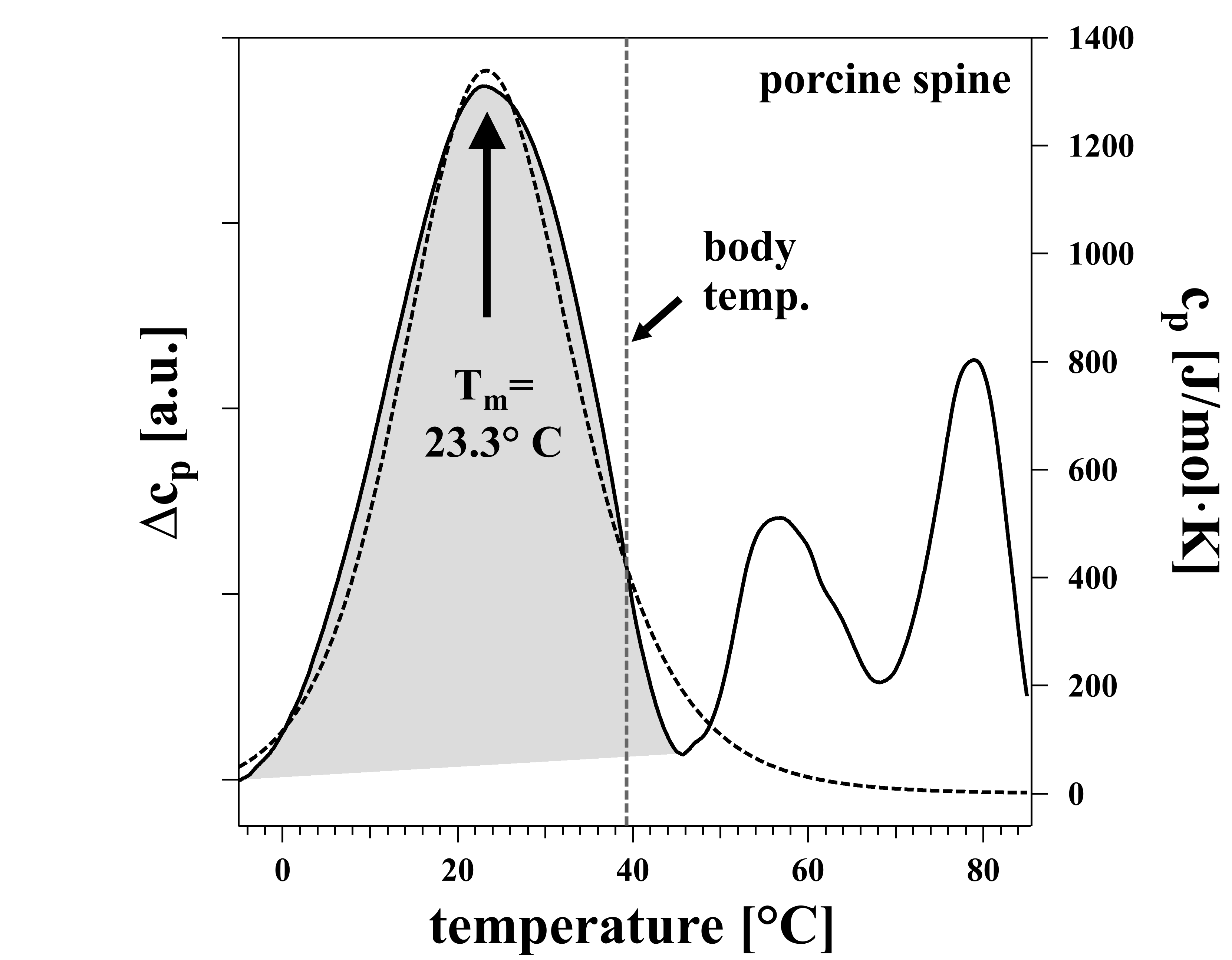}
\caption{Melting profile of native membranes from porcine spine. The gray-shaded area represents the lipid melting peak with a maximum at 23.3$^\circ$C. The peaks above body temperature show the protein unfolding peaks. The dashed profile has been calculated using eq. (\ref{eq:melting_03}) assuming $\Delta H_0=35$ kJ\slash mol and a cooperative unit size of $n=3.2$. Adapted from \cite{Wang2018}.}
\label{figure1}
\end{figure}

Fig.\ref{figure1} shows the the heat capacity profile of native membranes from porcine spine (from \cite{Wang2018}). One finds a pronounce peak representing lipid melting below body temperature shaded in grey. Further peaks above body temperature represent protein unfolding peaks which are irreversible in repeated temperature scans. The body temperature of pigs is around 39$^\circ$C.

\begin{figure*}
\centering
\includegraphics[width=15cm]{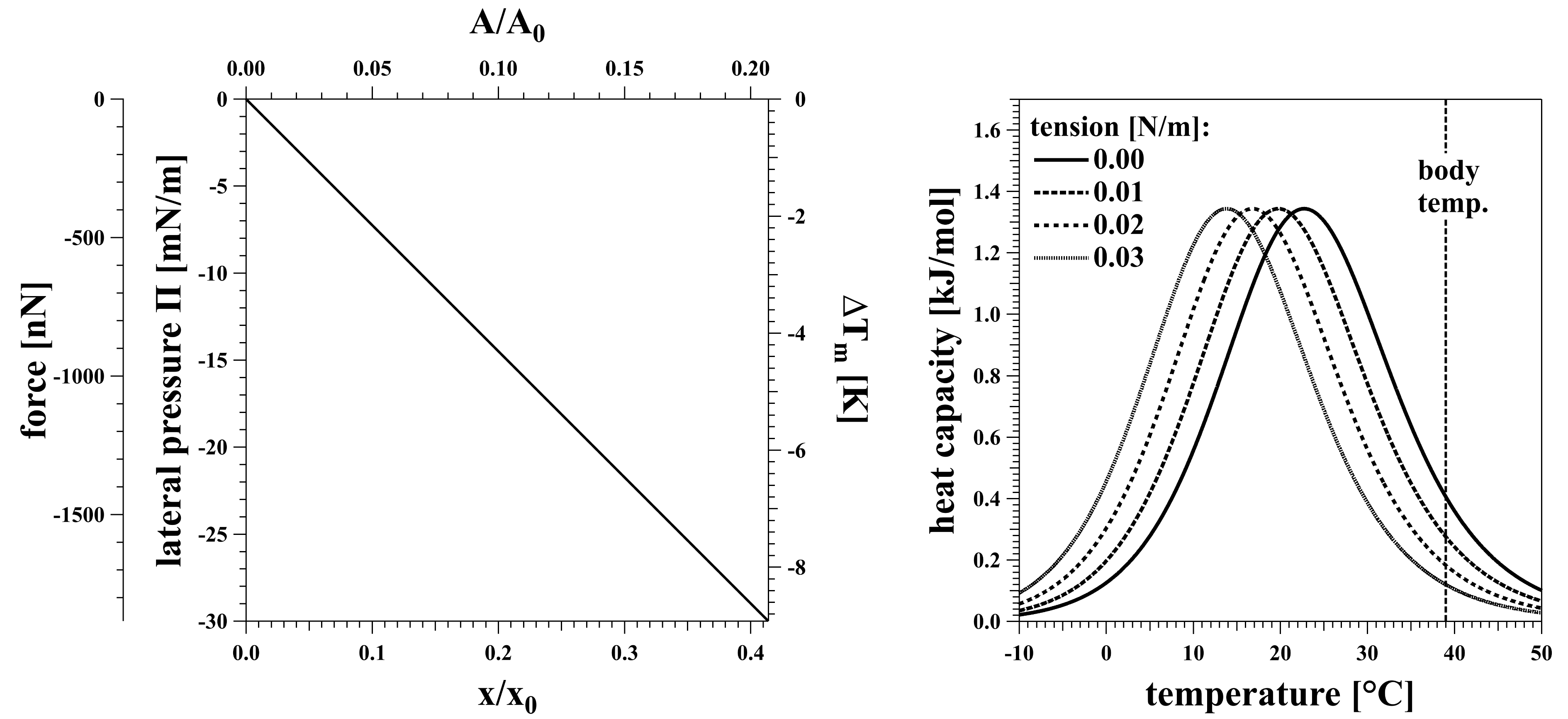}
\caption{\label{figure2}Left: Change in lateral tension (equal to -$\Pi$), force in an axon with a radius of $r_0=10 \mu$m, in melting temperature of a cylindrical neuron as a function of relative length changes, $x/x_0$. A lateral compressibility of 6.9 m\slash N for the liquid membrane is assumed (value for DPPC, \cite{Heimburg1998}). Right: Heat capacity profiles as a function of tension of a membrane with $\Delta H=35$ kJ\slash mol, a relative area change of $\Delta A/A_0=0.246$, a melting temperature of $23.3^\circ$C in the unperturbed axon and a cooperative unit size of $n=3.2$.}
\end{figure*}

The melting temperature $T_{m,0}$ of a membrane at a lateral tension of $\Pi=0$ is given by \cite{Heimburg2007a}
\begin{equation}
T_{m,0}=\frac{\Delta H_0}{\Delta S_0}\quad\rightarrow S_0=\frac{\Delta H_0}{T_{m,0}} \;.
\label{eq:melting_01}
\end{equation}

For DPPC membranes, $\Delta H = 35$ kJ\slash mol, and $T_{m,0}=314.2$ K (=41.05 $^\circ$C). In a nerve membrane, we use the measured melting temperature of 23.3$^\circ$C but further assume the same enthalpy of melting as for DPPC. If the tension is not equal to zero, the enthalpy change is given by $\Delta H=\Delta H_0+\Pi \Delta A_0$ instead. Thus, the dependence of the melting temperature on lateral tension is given by
\begin{equation}
T_{m}=\frac{\Delta H_0+\Pi\Delta A_0}{\Delta S_0}\quad\rightarrow \Delta T_m=\frac{\Pi \Delta A_0}{\Delta H_0}\cdot T_{m,0} \;.
\label{eq:melting_02}
\end{equation}
If $\Pi$ is negative (tension), the change in the transition temperature is negative, $\Delta T_m<0$.

The molar area of a DPPC membrane in the solid state at 25$^\circ$C is $A_0^{solid}$ =142728 m$^2$\slash mol, and the relative area change $\Delta A_0/A_0^{solid}=0.246$ \cite{Heimburg1998}. Thus, $\Delta A_0=35111$ m$^2$\slash mol. Assuming that the enthalpy and area changes of a biological membrane are similar to those of DPPC and that $T_{m,0}=23.3^\circ$C, we find for a lateral tension of $\Pi=-0.02$N\slash m that $\Delta T_m=-5.95$ K (see Fig. \ref{figure2}, left).
Using the above parameters, we can calculate the melting profile as a function of temperature using
\begin{eqnarray}
\label{eq:melting_03}
\Delta c_p&=&\frac{K}{(1+K)^2}\frac{n\Delta H^2}{RT^2} \;,\\
\mbox{where} && K=\exp{\left(-\frac{n(\Delta H-T\Delta S_0)}{RT}\right)} \;,\nonumber
\end{eqnarray}
which was derived in \cite{Heimburg2007a}. The cooperative unit size $n$ indicates how many lipids on average undergo a transition simultaneously. It determines the width of the melting profile. The dashed line in Fig. \ref{figure1} has been calculated using the above equation using for $\Delta H=35$ kJ\slash mol, $T_{m,0}=23.3^\circ$C and $n=3.2$. We can now calculate the melting profiles as a function of membrane tension. The result is shown in Fig. \ref{figure2}, right. We see how the melting profile shifts away from physiological temperature towards lower temperatures when the membrane tension assumes larger values (more negative lateral pressures).


\subsection{Soliton amplitudes as a function of energy}
\label{solitonamplitudesasafunctionofenergy}

At physiological temperature, the nerve membrane is in its fluid state. However, perturbations can move it locally through the phase transition into the solid state. This local excitation (a region of ordered membrane) can propagate along the nerve axon with a velocity $v$ that is somewhat smaller than the sound velocity within the membrane, $c_0$ \cite{Heimburg2005c}. The velocity is of order 100 m\slash s, i.e., of similar order than the velocity of nerve pulses in motor neurons. The propagating pulse is called a `soliton'. It displays all alterations in membrane properties that are associated to the melting transition, among those thickness changes, length changes, reversible heat exchange, and voltage changes for charged or polarized membranes \cite{Heimburg2005c}. Since all of these changes have been measured experimentally, it seems plausible to assume that the soliton is identical to the nerve pulse. A soliton is a special case of sound propagation. However, it consists of a localized pulse rather than of a sinusoidal wave due to the presence of the melting transition.

\begin{figure*}[ht]
\centering
\includegraphics[width=12cm]{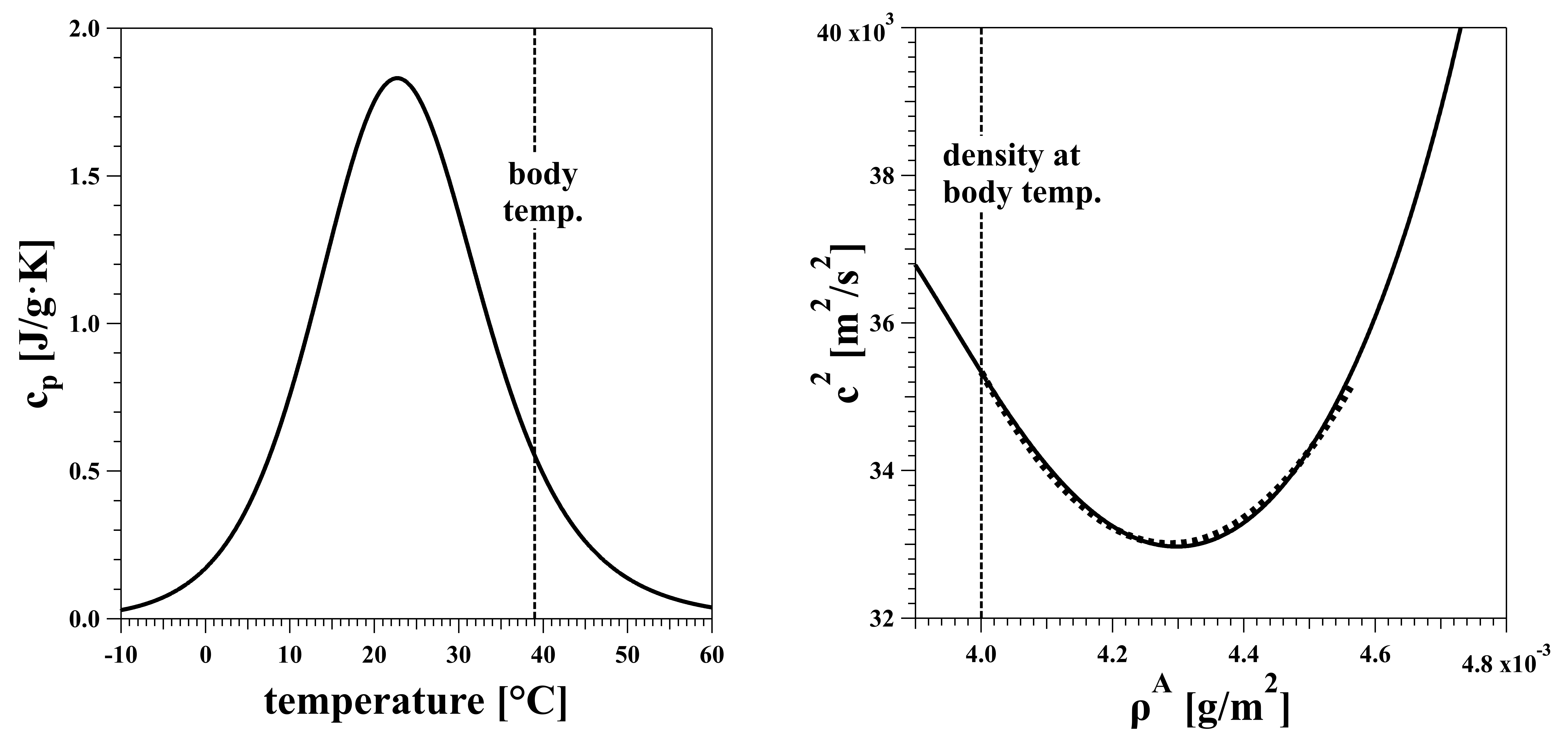}
\caption{\label{figure4}The melting transition in porcine spine nerves. Left: The theoretical heat capacity profile from Fig. \ref{figure1}. Body temperature is indicated by the dashed line. Right: Square of the low frequency sound velocity as a function of area density (solid curve). The vertical dashed line indicates the area density at body temperature. The dashed curve is the least-square fit of $c^2$ to a second order polynomial. }
\end{figure*}

The soliton describing the nerve pulse is given by the solution of the following differential equation derived in \cite{Heimburg2005c}
\begin{equation}
\frac{\partial^2 \rho^A}{\partial t^2}=\frac{\partial}{\partial x} \left(\left(c_0^2 + p\Delta \rho^A + q (\Delta \rho^A)^2\right)\frac{\partial \rho^A}{\partial x}\right)-h\frac{\partial^4 \rho^A}{\partial t^4} \;,
\label{eq:soliton_01}
\end{equation}
where $x$ is the coordinate along the nerve axon, and $t$ is time.
This equation possesses an analytical solution which is given in \cite{Lautrup2011}. The parameters in the above equation are $c_0$, the 2D sound velocity in the membrane at body temperature (around 170 m\slash s), $p$ and $q$, which are coefficients of a Taylor expansion of the sound velocity as a function of density close to the transition, and the dispersion parameter $h$ describing the frequency dependence of the sound velocity.

In the following we use the fit to the melting profile shown in Fig. \ref{figure1} as the basis for our calculations (see Fig. \ref{figure4}). Further, we assume that the enthalpy of melting as well as the compressibility of the solid and liquid membrane state and their temperature dependence are similar to those of DPPC. The respective parameters are summarized in \cite{Heimburg1998}. Fig. \ref{figure4} (right) shows the square of the speed of sound calculated from the heat capacity following the procedure shown in \cite{Heimburg1998} as a function of the area-density of the membrane. One can see that this function displays a minimum at a membrane density slightly higher than the density at body temperature. This minimum corresponds to the position of the melting transition. The dashed curve indicates the least square fit to $c^2=c_0^2 + p\Delta \rho^A + q (\Delta \rho^A)^2$, where $c_0^2=188$ m\slash s (sound velocity at body temperature), $\rho_0^A=0.0040$ g\slash m$^2$ (area density at body temperature), $p=-1.8291 \cdot c_0^2/\rho_0^A$, and $q=12.7193 \cdot c_0^2/(\rho_0^A)^2$. The vertical dashed line indicates the area density of the membrane at body temperature.

\begin{figure*}[ht]
\centering
\includegraphics[width=17cm]{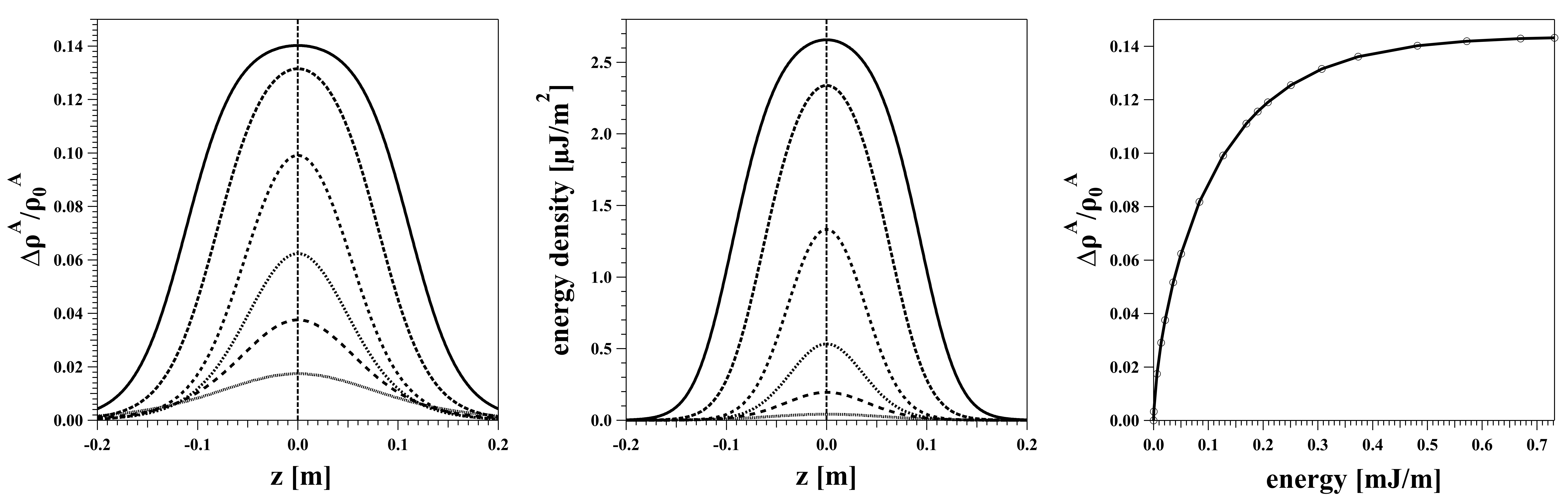}
\caption{\label{figure5}Solitons in porcine spine nerves. Left: The change in area density for different initial velocities. Center: The corresponding energy density profiles. Right: The soliton amplitude as a function of the energy density integrated over the soliton profile. For increasing energies, the amplitudes approach a maximum. }
\end{figure*}

Let us introduce a coordinate system that is moving with the velocity $v$ of the soliton. In this moving frame, the spatial coordinate is $z=x-vt$. With the above parameters on can determine the density profiles of the solitons as a function of z for different pulse velocities, which are given in Fig. \ref{figure5} (left). The energy density of the soliton is given by \cite{Heimburg2005c}
\begin{equation}
e=\frac{c_0^2}{\rho_0^A}(\Delta \rho^A)^2+\frac{p}{3\rho_0^A}(\Delta \rho^A)^3+\frac{q}{6\rho_0^A}(\Delta \rho^A)^4 \;.
\label{eq:soliton_01}
\end{equation}
It is is given in the center panel of Fig. \ref{figure5} for different velocities. The integral of the energy density is the total energy of the soliton. It carries the units J\slash m because in order to obtain absolute numbers it has to be multiplied by the circumference of the axon. Axons with larger diameter carry a large energy. In the right-hand panel of Fig. \ref{figure5}, the amplitude of the solitons is plotted as a function of the total energy of the soliton. One can recognize that upon increasing energy, the amplitude of the solitons increases until it reaches a maximum value around $(\Delta \rho/\rho_0)_{max}=0.144$. The energy of the soliton has to be provided from the outside by a stimulus in any variable that alters the membrane state.


\subsection{Free energy of membrane excitation}
\label{freeenergyofmembraneexcitation}

In the soliton theory, the nerve membrane is transiently shifted though its melting transition following an external stimulus provided by a local free energy change. This stimulus can be anything that renders a membrane more solid, i.e., an increase in lateral pressure \cite{Heimburg2005c}, a local cooling of the membrane \cite{Kobatake1971}, or a lowering of pH \cite{Muzic2019}. Thus, a decrease in lateral pressure (or an increase in tension) will act against soliton formation because it renders the membrane more fluid.

\begin{figure*}[ht]
\centering
\includegraphics[width=15cm]{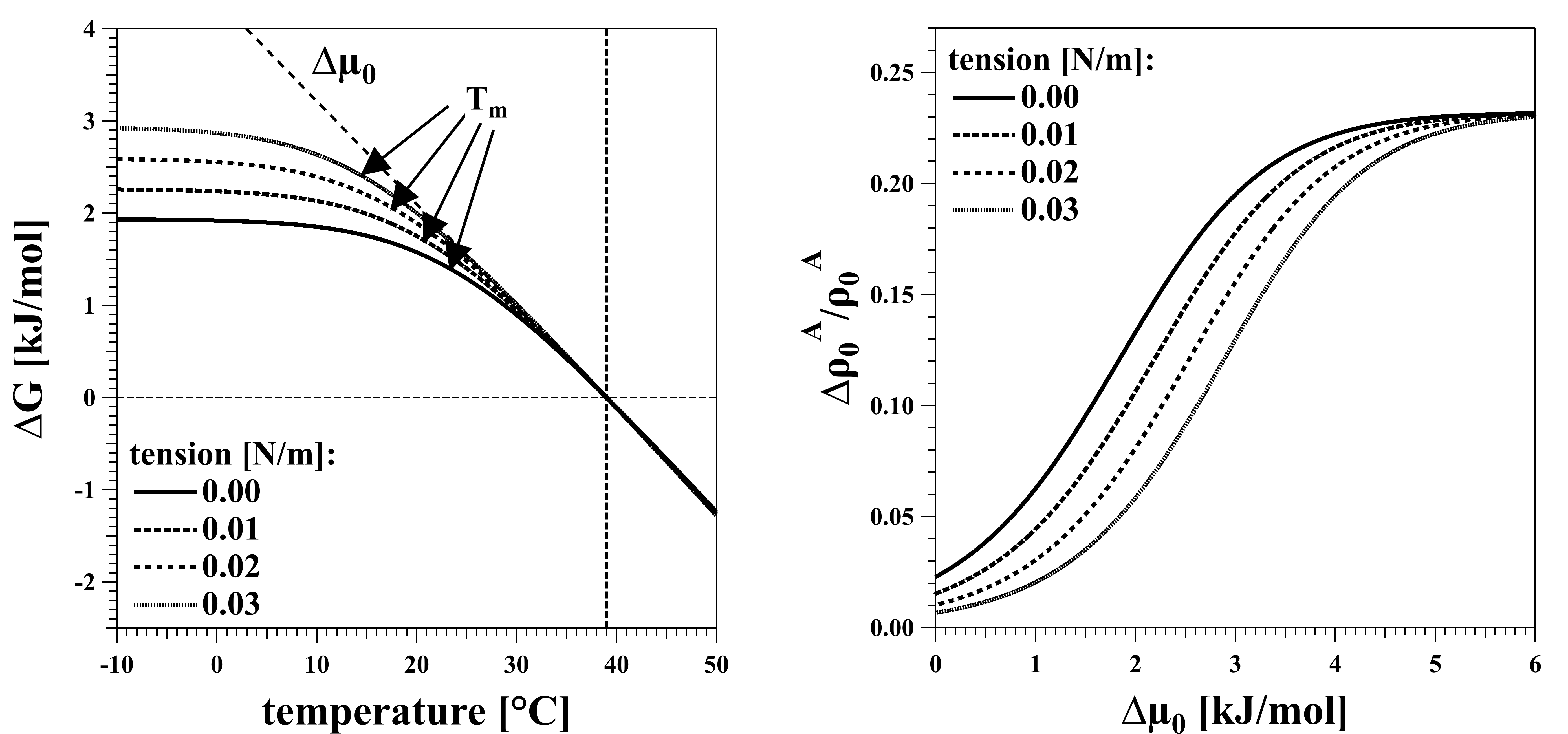}
\caption{\label{figure3}Left: The free energy difference between solid and liquid phase of the membrane relative to the free energy at 39$^\circ$C (the body temperature of pigs) for different tensions (negative lateral pressures). The dashed line represents the chemical potential difference between gel und fluid state as a function of temperature relative to the chemical potential at 39$^\circ$C. Right: The change in membrane area density as a function of the chemical potential difference for different lateral tensions. }
\end{figure*}

\begin{figure}[htbp]
\centering
\includegraphics[width=8cm]{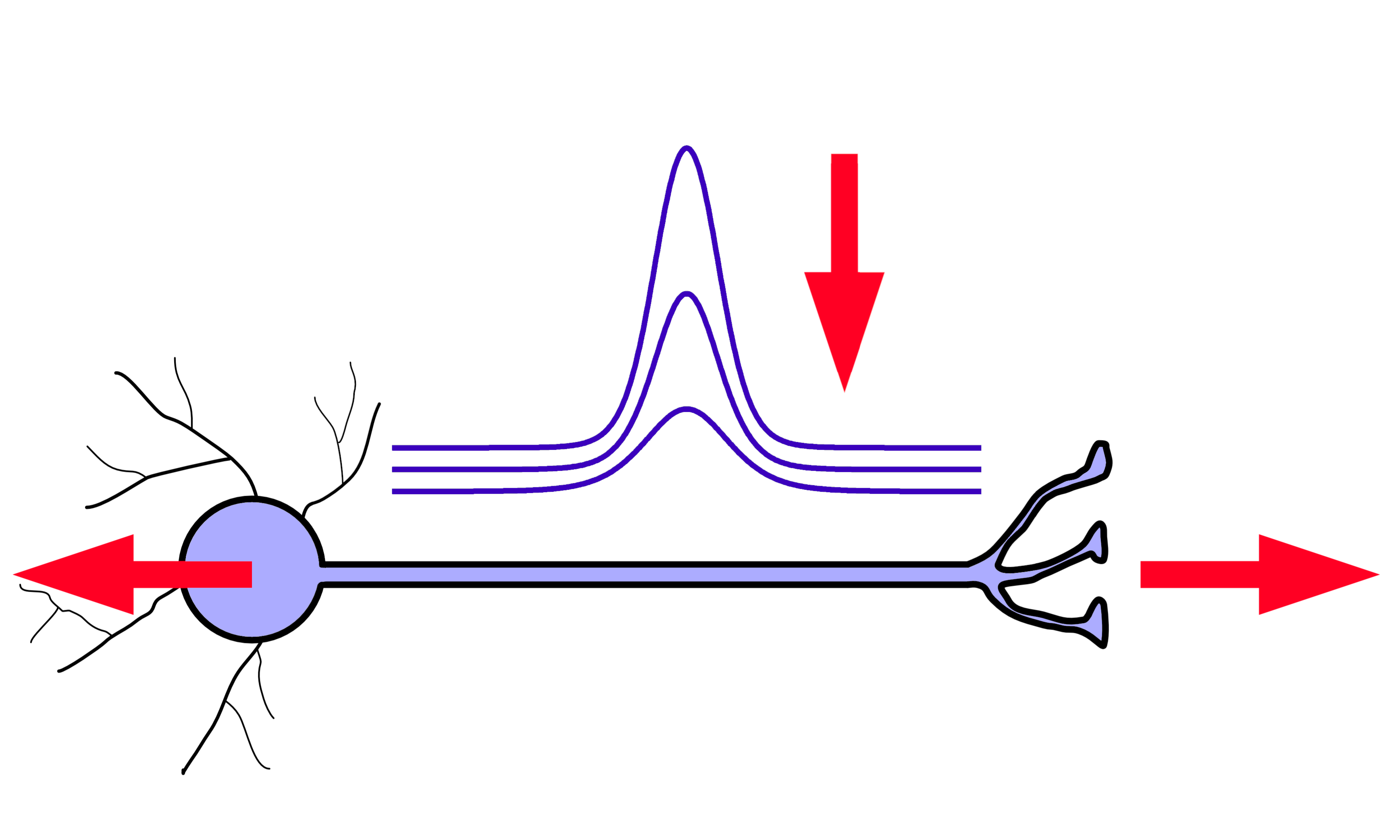}
\caption{Schematic drawing of a nerve under stretch and the resulting reduction in nerve pulse amplitude for a constant stimulus.}
\label{figure6}
\end{figure}

The chemical potential difference between the solid and the liquid membrane state is given by
\begin{equation}
\Delta \mu_0=\Delta H-T\cdot \Delta S_0 \;,
\label{eq:chempot_01}
\end{equation}
where $\Delta H=\Delta H_0+\Pi\Delta A_0$ and $\Delta S_0=\Delta H_0/T_{m,0}$ as in eq. (\ref{eq:melting_02}). Here, both $\Delta H_0$ (the enthalpy at zero tension, i.e. $\Pi=0$), $\Delta A_0$ and $\Delta S_0$ are constants.
 The chemical potential difference in eq. (\ref{eq:chempot_01}) displays a linear dependence on temperature with a slope of $-\Delta S_0$. The temperature at which $\Delta \mu_0=0$ (the melting temperature) shifts as a function of tension. The dashed line in Fig. \ref{figure3} (left) shows the difference of the chemical potential relative to the chemical potential at the body temperature of pigs (39$^\circ$C), i.e., $\mu_0(T)-\mu_0(39^\circ$C).
The free energy difference of a membrane at temperature $T$ relative to the free energy of the solid membrane is
\begin{equation}
\Delta G(T)= \frac{K}{1+K}\Delta \mu_0
\label{eq:chempot_04}
\end{equation}
where $K$ was defined in eq. (\ref{eq:melting_03}), and $f=K/(1+K)$ is the fluid faction of the membrane. The curved lines in Fig. \ref{figure3} (left) show the free energy difference relative to the free energy at body temperature (39$^\circ$C), $\Delta G_0(T)-G_0(39^\circ$C). It depends on the lateral tension, $-\Pi$. If we consider a membrane at body temperature, the free energy required to render the membrane solid depends and the lateral tension. If we relate this to the free energy to excite a soliton, we find that the higher the tension, the more free energy is required the render the membrane solid. Or, at constant stimulus, the nerve axon the soliton amplitude decreases as can be seen in Fig. \ref{figure3} (right). For constant free energy change, the density change becomes smaller with increasing tension. This will be reflected in the amplitude of nerve solitons, as schematically shown in Fig. \ref{figure6}.

The lowering in the transition temperature $T_m$ induced by membrane tension resembles the effect of anesthetics on membranes. The anesthetic concentration is also related to the chemical potential and the free energy difference between fluid and solid phase of the nerve membrane. In \cite{Wang2018}, we discussed how this effects the stimulation threshold for nerve excitation. The line of argument is similar to the ones cause by tension described here.


\section{Discussion}
\label{discussion}

Here, we provide a theoretical study on the excitability of nerve membranes while stretching the nerve within the context of the soliton theory for nerve pulse propagation and excitation. In this theory, the soliton consists of a compressional pulse, i.e., it is profoundly coupled to the mechanical properties of the nerve membrane, which are a function of membrane tension. We found that nerve stretching is expected to inhibit the generation of a nerve pulse. Stretching of nerves could originate from the bending of a knee or an elbow. It has been reported that the H reflex in the back of the knee depends on the bending of the knee. Similarly, the T-reflex (patella reflex) depends on bending the knee. Thus, the above considerations may play a role in physiology. In a living organism, stretching is probably on a small scale ( a few percent in length). However, we show that even such small stretches can have a significant effect on the melting transition in nerve membranes, which influences the nerve excitability.

It is known that the nerve pulse has mechanical components. Nerves contract during activity \cite{Wilke1912a, Wilke1912b, Tasaki1982a, Tasaki1989}. This suggests that stretching the nerve might influence pulse propagation and generation. Nerves also change their thickness \cite{Iwasa1980a, Iwasa1980b, Tasaki1982b, Tasaki1989, Kim2007, GonzalezPerez2016, Ling2020} and they can be excited mechanically \cite{Tigerstedt1880, Schmitz1938, Yamada1961, Hashimoto1989}. Further, nerves can be stimulated by ultrasound, which lead to the emerging field of focussed ultrasound neurostimulation (FUN) \cite{Tufail2011, Mueller2014, Jerusalem2019}. With this method brain regions can be stimulated in a noninvasive manner. For instance, FUN of the brain of mice can stimulate the contraction of legs and the tail \cite{Kamimura2016}. Thus, there exists clear experimental evidence that nerve pulse propagation is not only accompanied by mechanical changes in the nerve axon, but that nerves can be excited by mechanical means. This explains the appropriateness of the soliton theory for treating such phenomena, for which the Hodgkin-Huxley theory and derivatives of it have no language.

The soliton theory is closely related to the theory of sound in membranes. It is of mechanical and thermodynamic nature. It does not consider individual macromolecules but rather the macroscopic physical properties of the membrane as a whole. It seems plausible to apply the soliton theory to the problem of nerve stretching. In the soliton picture, the nerve pulse consists of a segment of gel membrane traveling in a fluid environment. The area density at the maximum of the soliton is about 20\% higher as in the unperturbed membrane state. Thus, the soliton is accompanied by a shortening of the nerve as observed in experiments (see above).
Nerve excitation requires the provision of free energy to the membrane in order to move it from a fluid to a gel state. Lateral tension decreases the melting temperature of a membrane because the liquid state of the membrane displays a larger area than the solid state. The free energy necessary to move the membrane into a gel state increases. As a consequence, it becomes more difficult to excite a nerve. Generally we predict that nerves become less excitable when they are stretched.

It has been widely reported that nerve stretching influences the generation of action potentials and generally decreases the amplitude of compound action potentials, e.g., in hamster sciatic nerve \cite[]{Stecker2011}. Rapid conduction loss has been observed at only 6.2\% nerve strain \cite{Wall1992}.\cite[]{Ochs2000} argued that the decrease in action potential amplitude could be a side effect of stretching by reducing the supply of the axons leading to ischemia and anoxia after a long stretch. This might be true for some experiments, but is unlikely for experiments during which the onset of the decrease in excitability is fast and reversible. Wall \cite[]{Wall1992} found that stretching a nerve to a strain of 6\% longitudinally in rabbit tibial nerve produced a 70\% reduction in the nerve action potential which returned to normal after releasing the tension. Only after 12\% sketch, the recovery was incomplete. This seems to indicate that the nerve at 6\% stretching was not injured while it was damaged after a stretch of 12\%. Other studies have found that nerves can be stretched considerable more without damage. Peripheral nerves are said to tolerate 15\% elongation before conduction is impaired \cite{Brown1989}.
Single nerve fibers from frog have been reported to tolerate elongations of 30\% \cite{Schneider1952} and 100\% in nerves from a slug \cite{Turner1951}.\\
Lengthening movements and static stretching of the human triceps have been investigated over many decades. Budini and Tilp \cite[]{Budini2016} find that the H-reflex decreases immediately as static stretching is applied and in proportion to the stretch degree. Within 2 s after a single passive dorsiflexion movement, H-reflex is strongly inhibited. This implies that stretching leads to a decrease of the H-reflex, and that this is an effect with a fast onset which is reversible. Stretch over longer time-scales or repetitive stretching leads to slow recovery, possibly due to the damage of nerves. Guissard and collaborators \cite[]{Guissard2001, Guissard2006} showed that motorneuron excitability decreased during fast muscle stretch. This inhibition had been tested in the soleus muscle by recording the Hoffmann reflex (H reflex) and the tendon reflex (T reflex) which were both found to be reduced during fast muscle stretching. Here, we argue that the reduction in action potential amplitude is a natural consequence of the effect of stretching on the physical state of the nerve membrane, which can well occur immediately after stretching without that the nerve is damaged.

Turner \cite{Turner1951} found that the pulse velocity was not significantly altered upon stretching, which is also the result of \cite{Goldman1964} who saw a constant conduction velocity in earthworm nerves stretched up to 450\%. It is worthwhile noting that the Hodgkin-Huxley model for nerve pulse conduction \cite{Hodgkin1952b} predicts that the conduction velocity $v$ depends on the radius $r$ of an axon such that $v\propto \sqrt{r}$. Thus, one expects a significant reduction in conduction velocity upon stretch in the HH model, which is inconsistent with the above findings. In contrast, the soliton theory predicts a constant propagation velocity for constant amplitude of the action potential.

In \cite{Heimburg2007c} we discussed the effect of anesthetics on the melting point in membranes. The shift of the transition temperature critical anesthetic dose, ED$_{50}$, at which 50\% of all individuals are anesthetized, was found to be of the order of -0.6$^\circ$ in tadpoles, and probably somewhat higher (around -1$^\circ$) in humans. Here we argue that like the effect of anesthetics, stretch lowers the melting transition of the nerve membrane and leads to a lower amplitude of action potentials at constant stimulus - or or higher stimulus required for maximum amplitude. According to Fig. \ref{figure1}, this corresponds to a stretch of the membrane by $\Delta x/x_0=0.028-0.046$, i.e., 2.8--4.6\%. The force require to stretch nerve with a radius of 10 \textmu m by this amount is 127--211 nN. A stretch by 10\% would already correspond to 2.2--3.6 times the critical anesthetic dose. Stecker et al \cite{Stecker2011} in fact found for the sciatic nerve of hamster that anesthetics interferes with the results from a stretching experiment.


\section{Summary}
\label{summary}

Summarizing, we predict that the stretching of nerves will reduce their excitability in a manner reminiscent of the action of anesthetics. At constant stimulus, stretching will result in a decrease of the action potential amplitude du to its effect on the melting transition in membranes. While there have been reports attributing the effect of stretching on the damage of nerves (which is a possibility for large and long stretch), there is quite some experimental evidence that suggests that nerve stretching can have a fast and reversible effect on reducing nerve activity and excitability. This supports the theoretical results from the soliton theory presented here. Since nerves are likely to change their length and tension during exercise, this implies that nerve activity may depend on the contraction of muscles and can be variable during exercise due to a feedback between muscle contraction and the elongation of nerves.





\small{

}

\end{document}